\documentstyle [preprint,aps,eqsecnum,epsfig,fixes,floats,lscape]{revtex}
\makeatletter           % Switch to floating figures
\@floatstrue
\def\figure{\let\@capwidth\columnwidth\@float{figure}}
\let\endfigure\end@float
\@namedef{figure*}{\let\@capwidth\textwidth\@dblfloat{figure}}
\@namedef{endfigure*}{\end@dblfloat}
\def\table{\let\@capwidth\columnwidth\@float{table}}
\let\endtable\end@float
\@namedef{table*}{\let\@capwidth\textwidth\@dblfloat{table}}
\@namedef{endtable*}{\end@dblfloat}
\makeatother
\def\ibid{$\it {ibid}.$}
\def\Tc{T_{\rm c}}
\def\rc{r_{\rm c}}
\def\kB{k_{\rm B}}
\def\etal{{\it et al.}}

\def\x{{\bf x}}

\def\lat{{\rm lat}}

\def\phiphic{\langle\phi^2\rangle_{\rm c}}

\def\c{{\rm c}}
\def\msb{{\overline{\rm MS}}}
\def\MSbar{$\msb$}
\def\DphLat{\Delta\langle\Phi^2\rangle_{\rm L}}
\def\DphCont{\Delta\langle\phi^2\rangle_{\rm c}}

\def\i{{\bf i}}

\def\blparen{\mbox{\boldmath$($}}
\def\brparen{\mbox{\boldmath$)$}}

\def\<{\langle}
\def\>{\rangle}

\def\R3dk{{\int_{{\cal R}^3}{d^3k\over (2 \pi)^3}}}

\def\asc{a_{\rm sc}}
\def\lat{{\rm lat}}

\tightenlines
\begin {document}

%preprint {UW/PT-01-07}

%\title{
%   Monte Carlo Simulation of O(1) and O(4) $\phi^4$ theory at
%  critical temperature in three dimensions
%}
\title{
  Monte Carlo studies of three-dimensional O(1) and O(4) \boldmath{$\phi^4$}
  theory related to BEC phase transition temperatures
}

\author {Xuepeng Sun}

\address
    {%
    Department of Physics,
    University of Virginia,
    P.O. Box 400714,
    Charlottesville, VA 22904--4714
    }%
\date {\today}

\maketitle
\begin {abstract}
The phase transition temperature for the Bose-Einstein condensation (BEC)
of weakly-interacting Bose gases in three dimensions is known to be related
to certain non-universal properties of the phase transition of
three-dimensional O(2) symmetric $\phi^4$ theory.  These properties
have been measured previously in Monte Carlo lattice simulations.
They have also been approximated analytically, with moderate success,
by large $N$ approximations to O($N$) symmetric $\phi^4$
theory. To begin investigating the
region of validity of the large $N$ approximation in
this application, I have applied the same Monte Carlo
technique developed for the O(2) model (Ref.\ \cite{O2sim})
to O(1) and O(4) theories. My results indicate that there might 
exist some theoretically unanticipated systematic errors in the 
extrapolation of the continuum
value from lattice Monte Carlo results.
% which have caused about
%10\% systematic uncertainty in my final results. 
The final results show that the difference between simulations
and NLO large $N$ calculations does not improve significantly
from $N=2$ to $N=4$. This suggests one would need to 
simulate yet larger $N$'s to see true large $N$ scaling
of the difference. Quite unexpectedly (and presumably 
accidentally), my Monte Carlo result for
 $N=1$ seems to give the best agreement with the large $N$
approximation among the three cases.
\end {abstract}

%%%%%%%%%%%%%%%%%%%%%%%%%%%%%%%%%%%%%%%%%%%%%%%%%%

\section { Introduction }

The computation of the phase transition temperature $\Tc$ for dilute or
weakly-interacting bose gases has attracted considerable interest. Due to the non-perturbative nature of long-distance
fluctuations at the second-order phase transition at large distance,
the calculation of corrections to the ideal gas formula for $\Tc$ is
non-trivial.
In the dilute or weakly-interacting limit, the
correction $\Delta \Tc \equiv T_c -T_0$ to the ideal
gas result $T_0$ for a homogeneous gas can be parametrized as%
\footnote{
   For a clean argument that the first correction is linear in
   $\asc$, see Ref.\ \cite{BBHLV}.  For a discussion of higher-order
   corrections, see Refs.\ \cite{AMT,log2}.
}
\begin {equation}
   {\Delta \Tc \over T_0}
   = c \asc n^{1/3}
        + O\blparen \asc^2 n ^ {2/3} \ln(\asc n^{1/3}) \brparen , 
\label{eq:infA1}
\end {equation}
where $a_{sc}$ is the scattering length of the 2-particle interaction, $n$
is the number density of the homogeneous gas, $c$ is a numerical constant,
and $O(\cdots)$ shows the parametric size of higher-order corrections.
Baym \etal\ \cite{BBHLV} have shown that the computation of $c$ can be 
reduced to a problem in three dimensional O($2$) $\phi^4$ field theory.
In general, O($N$) $\phi^4$ field theory is described by the continuum
action
\begin {equation}
  S_{\rm cont} = \int d^3x \left[
     {1\over2}(\nabla\phi)^2
     +{1\over2}r\phi^2
     +{u \over 4!}\left(\phi^2\right )^2
  \right] ,
\label{eq:3Dact}
\end {equation}
where $\phi=(\phi_1, \phi_2, \cdots, \phi_N)$ is an $N$-component
real field.
I will focus exclusively on the case where
$r$ has been adjusted to
be at the order/disorder phase transition for this theory
for a given value of the quartic coupling $u$.  The relationship to
$\Tc$ for Bose-Einstein condensation found by Baym \etal\ is that the
constant $c$ in (\ref{eq:infA1}) is%
\footnote{
   This is given separately in Ref.\ \cite{BBHLV} as
   $\Delta \Tc / T_0 = - 2 m \kB T_0 \,\Delta\phiphic / 3\hbar^2 n$
   and the identification of $u$ as $96\pi^2\asc/\lambda^2$.
}
\begin {equation}
   c = - {128 \pi^3 \over \left[\zeta({3\over2})\right]^{4/3}} \,
       {\Delta\phiphic\over u} \,,
\label {eq:c}
\end {equation}
where $\phi^2 \equiv \phi_1^2 + \phi_2^2 + \cdots \phi_N^2$ , $N=2$ and 
\begin {equation}
   \Delta\phiphic \equiv \left[\phiphic\right]_u - \left[\phiphic\right]_0
\end {equation}
represents the difference between the critical-point value of 
$\langle\phi^2\rangle$ for the cases of (i) $u$ non-zero and (ii)
the ideal gas $u{=}0$.
Thus the computation of the first correction to $T_c$ due to
interactions is related to the evaluation of $\Delta\phiphic$
in three-dimensional $\phi^4$ theory.

Having tuned $r$ to the phase transition, $u$ is then the single remaining
parameter of the three-dimensional continuum $\phi^4$ theory (\ref{eq:3Dact}).
The dependence of (ultraviolet-convergent) quantities on $u$ is
determined by simple dimensional analysis, and $u$ has dimensions of inverse
length.  The $\Delta\phiphic/u$ in (\ref{eq:c}) is dimensionless and so
is a number independent of $u$ in the continuum theory.  Monte Carlo
simulations of this quantity in O(2) theory have given
$c=1.29\pm 0.05$ \cite{KPS} and $c=1.32\pm 0.02$ \cite{O2sim}.

One of the few moderately successful attempts to approximate this result
with an analytic calculation has been through the use of the large $N$
approximation.  (But see also the 4th-order linear $\delta$ expansion
results of Refs.\ \cite{delta}.  For a brief comparison
of the results of a wide spread of attempts
to estimate $c$, see the introduction to Ref.\ \cite{O2short} and also
Ref.\ \cite{c}.).
The procedure is to calculate $\Delta\phiphic/u$ for
O($N$) theory in the limit that $N$ is arbitrarily large, and then
substitute the actual value $N{=}2$ of interest into the result.
The large $N$ result was originally computed at leading order in $1/N$ by Baym,
Blaizot, and Zinn-Justin \cite{BBZ} and extended to next-to-leading
order (NLO) in $1/N$ by Arnold and Tom\'{a}\v{s}ik \cite{AT},
giving
\begin{equation}
  {\Delta\phiphic \over u}
  = -{N\over 96\pi^2}\left[1-{0.527198\over N} + O(N^{-2}) \right] .
\label{eq:dphiN}
\end{equation}
Setting $N=2$, one obtains $c \sim 1.71$, which is roughly 30\% higher
than the results obtained by Monte Carlo simulation of O(2) theory.
Considering that $N=2$ has been treated as large in this approximation,
the result is fairly encouraging.

The goal of the present work is to further explore the applicability of
the large $N$ result (\ref{eq:dphiN}) by testing it for other values of
$N$.  I have applied to other O($N$) models the same techniques used
in Ref.\ \cite{O2sim} to simulate the O(2) model.
In this paper I report the measurement of $\Delta \phiphic /u$ for the
O(1) and O(4) theories.  The final results, compared to the large $N$
approximations of (\ref{eq:dphiN}), are given in table
\ref{tab:summary}.

As a byproduct of the analysis, I also report 
the measurement of the critical value $\rc$ of $r$.
The coefficient $r$ requires ultraviolet renormalization and so
is convention dependent.
In Table \ref{tab:summary}, I report the dimensionless continuum values of
$r_c / u^2$ with $r_c$ defined by dimensional regularization and 
modified minimal subtraction (\MSbar) renormalization at a renormalization
scale $\bar\mu$ set to $u/3$.  Among other things, this quantity can
be related to the coefficient of the second-order ($\asc^2n^{2/3}$)
correction to $\Delta T_c$ \cite{AMT,AT2}.
One can convert to  other choices of the \MSbar\ renormalization
scale $\bar\mu$ by the (exact) identity
\begin {equation}
   {r^\msb(\bar\mu_1)\over u^2} = {r^\msb(\bar\mu_2)\over u^2}
      + {(N+2)\over18(4\pi)^2}\, \ln{\bar\mu_1\over\bar\mu_2} .
\label {eq:mubarconvert}
\end {equation}

\begin{table}
\begin{center}
\begin{tabular} {|l||l|l|r||l|}
\hline
\multicolumn{1}{|c||}{$N$} &\multicolumn{3}{|c||}{${\Delta\phiphic/ u}$}
                          &  $\rc/u^2(\bar\mu{=}u/3)$ \\
\cline{2-4} 
& \multicolumn{1}{|c|}{simulation} & 
\multicolumn{1}{|c|}{large $N$} &
\multicolumn{1}{|c||}{difference} &
\multicolumn{1}{|c|}{simulation}\\
\hline
1 & -0.000494(41)& -0.0004990 & -~~1(8) \% & 0.0015249(48)\\
\hline
2 (Ref.\ \cite {O2sim})
  & -0.001198(17) & -0.001554 & +30(2) \% & 0.0019201(21)\\
\hline 
4 & -0.00289(18) &-0.003665 & +27(8) \% & 0.002558(16) \\
\hline
\end{tabular}
\end{center}
\caption{
  Simulation results for $N$=1,2,4 and the NLO large $N$ results
  for $\Delta \phiphic /u$. 
  The difference column shows the percentage excess of the
  magnitude of the large $N$ approximation result
  over the magnitude of the simulation result.
  $N=2$ simulation results are quoted from Ref.\ \protect\cite{O2sim}.
}
\label{tab:summary}
\end{table}

\par
In section \ref{sec:basics}, I briefly review the algorithm and
the necessary background materials and formulas for improving the
extrapolations of the continuum and infinite-volume limits.
Most of the technical details can be found
in Ref.\ \cite {O2sim}. 
Section \ref{simresults}  gives the simulation results and analysis.
Section \ref {concls} is the conclusion from comparing the numerical
results with the NLO large $N$ calculation (\ref{eq:dphiN}). 

%%%%%%%%%%%%%%%%%%%%%%%%%%%%%%%%%%%%%%%%%%%%%%%%%%%%%%%%

\section {Lattice Action and Methods}
\label{sec:basics}

The action of the theory on the lattice is given by  
\begin{eqnarray}
  S_{lat} & = & a^3\sum_x\{{1\over 2}\Phi_\lat(-\nabla_\lat^2)\Phi_\lat
  +{1\over2}r_0\Phi_\lat^2+{u_0\over 4!}(\Phi_\lat^2)^2\} ,
\end{eqnarray}
where $a$ is the lattice size (not to be confused with the scattering
length $\asc$).
I will work on simple cubic lattices with cubic total volumes
and periodic boundary conditions.
As in Ref.\ \cite{O2sim} and as reviewed further below,
the bare lattice operators and couplings
($\Phi_\lat$, $\Phi_\lat^2$, $r_0$, $u_0$) are matched to 
the continuum, using results from
lattice perturbation theory to
improve the approach to the continuum limit for finite but small lattice
spacing.

I shall use the improved lattice Laplacian
\begin{equation} 
  \nabla^2\Phi(\x) = a^{-2} \sum_\i \Bigl[ {\textstyle
       -\frac{1}{12}\Phi(\x+2a\i)+\frac{4}{3}\Phi(\x+a\i)-\frac{5}{2}\Phi(\x)
       +\frac{4}{3}\Phi(\x-a\i)-\frac{1}{12}\Phi(\x-2a\i)
   } \Bigr] ,
\label{eq:laplacian}
\end{equation}
which (by itself) has $O(a^4)$ errors, rather than the standard unimproved
Laplacian
\begin {equation}
\nabla^2_{\rm U} \Phi(\x) = a^{-2}
   \sum_\i \Bigl[\Phi(\x+a\i)-2\Phi(\x)+\Phi(\x-a\i)\Bigr] 
\label {eq:lapU} ,
\end {equation}
which has $O(a^2)$ errors.
Readers interested
in simulation results with the unimproved kinetic
term should see Ref.\ \cite{thesis}. 
% Our experience is that using the improved Laplacian gives
% cleaner, less ambiguous extrapolations of the continuum limit, even though
% there remain other sources of $O(a^2)$ errors in our extrapolation of the
% continuum limit.

Because the only parameter of the continuum theory (\ref{eq:3Dact})
at its phase transition is $u$, the relevant distance scale for the physics
of interest is of order $1/u$ by dimensional analysis.  To
approximate the continuum infinite-volume limit, one needs lattices
with total linear size $L$ large compared to this scale and
lattice spacing $a$ small compared to this scale.
As a result, $ua$ is the relevant dimensionless expansion parameter for
perturbative matching calculations intended to eliminate
lattice spacing errors to some order in $a$.  The analogous dimensionless
parameter that will appear in the later discussions of the large volume
limit is $Lu$.

% ------------------------------------------------------------------------

\subsection {Matching to continuum parameters}

To account for the lattice spacing errors, I have adapted
the perturbative matching calculations to improve lattice
spacing errors given in Ref.\ \cite{O2sim}, where the details of the calculations can be found.  Here I will simply
collect the results from that reference.
The lattice action expressed in terms of continuum 
parameters is written in the form
\begin {eqnarray}
  S_{\rm lat} & \equiv & a^3 \sum_x \left \{
    {Z_\phi \over 2} (\nabla_{\rm lat} \phi)^2 + { Z_r\over 2} ( r+\delta r) \phi^2
+ { u + \delta u \over 4 !} ( \phi^2) ^2 \right \} .
\label {latact}
\end {eqnarray}
The continuum approximate value for $\Delta\phiphic$ is
theoretically expected to have the form
\begin {equation}
   \Delta\phiphic 
   = Z_r \langle\phi^2\rangle_\lat - \delta\phi^2 + O(a^2).
\label{eq:phi2renorm}
\end {equation} 
For the improved Laplacian, the matching calculations have been done 
to two-loop level ($2l$) for $Z_\phi$, $Z_r$, $\delta r$ and $\delta u$,
and three loops ($3l$) for $\delta\phi^2$, yielding
\begin{mathletters}
\begin {eqnarray}
   \delta\phi^2_{3l} &=&
     { {N \Sigma}\over4 \pi a} \,
     + {N(N+2)\over 6} \, {\Sigma \xi \over (4\pi)^2} \, u 
     - \frac{\xi}{4 \pi} \, N r a  
\nonumber \\ && \qquad
     + \left[ \left(N+2\over6\right)^2 \xi^2\Sigma 
       + {(N+2)\over 18} \left(
           C_4 - 3 \Sigma C_1 - \Sigma C_2 + \xi \log(a\bar\mu) \right)
        \right] {N u^2 a \over (4\pi)^3} , 
\label {dphi_ct}
\\
  \delta u_{2l} &=& {(N+8)\over 6} {\xi \over 4 \pi} u^2 a+ 
    \left[
       {(N^2+6N+20)\over 36}\left(\xi\over4\pi\right)^2
       - {(5N+22)\over9}\, {C_1\over(4\pi)^2}
    \right] u^3 a^2 ,
\\
  Z_{\phi,2l} &=& 1 +
    {(N+2)\over 18} \, {C_2\over (4\pi)^2} \, u^2 \, a^2 ,
\\
  Z_{r,2l} &=& 1 + {(N+2)\over 6} {\xi \over 4 \pi} ua  +
 \left[\left(N+2\over6\right)^2\left(\xi\over4\pi\right)^2
    - {(N+2)\over6} \, {C_1\over(4\pi)^2} \right] (ua)^2 ,
\\
  \delta r_{2l}  &=& -{(N+2)\over 6} {\Sigma\over 4 \pi} ua +
    {(N+2)\over18 (4\pi)^2} \left[
        C_3 + \ln\left(6\over\bar\mu\right) - 3 \Sigma \xi
    \right] (ua)^2 ,
\label{match_cts}
\end {eqnarray}
\end{mathletters}%
where, for the improved Laplacian (\ref{eq:laplacian}),
\begin {mathletters}%
\label {eq:consts}
\begin {eqnarray}
   \Sigma &\simeq& 2.75238391130752,
\\
   \xi &\simeq& -0.083647053040968 ,
\\
   C_1 &\simeq& 0.0550612 ,
\\
   C_2 &\simeq& 0.03344161 ,
\\
   C_3 &\simeq& -0.86147916 ,
\\ 
   C_4 &\simeq& 0.282  .
\end {eqnarray}%
\end {mathletters}%

As shown in Ref.\ \cite{O2sim}, the result of this 
improvement is that at fixed physical system size $Lu$, 
the lattice spacing error of $\Delta\phiphic /u$ should
be $O(a^2)$. However as will be shown in the following,
my simulation results indicate that there might still
exist some linear coefficients even after applying the
formula (\ref{eq:phi2renorm}). I assign an error to
my continuum extrapolations that covers both linear and 
quadratic extrapolations.

This has produced
about 10\% systematic error in the final value of
$\DphCont/u$. The extrapolation of $\rc/u^2$ on the other
hand has $O(a)$ error, since no improvement has been made.
Overall, the fitting formulas for the data taken at 
fixed $Lu$ are given as
\begin {mathletters}
\label{eq:Lufx1}
\begin{eqnarray}
   \left\{ {\Delta\phiphic \over u}\right\} _{Lu} 
   &=& B_1+B_2 (ua)^2 ,
\\
   \left \{ {r_c \over u^2} \right \}_{Lu} 
    & = & D_1 + D_2(ua), 
\label {eq:Lufx1_rc}
\end{eqnarray}
\end{mathletters}%
for a quadratic fit of $\DphCont/u$, and 
\begin {mathletters}
\label{eq:Lufx2}
\begin{eqnarray}
   \left\{ {\Delta\phiphic \over u}\right\} _{Lu} 
   &=& B_1+B'_2 (ua),
\label{eq:Lufx2_dphi}
\\
   \left \{ {r_c \over u^2} \right \}_{Lu} 
    & = & D_1 + D_2(ua), 
\label {eq:Lufx2_rc}
\end{eqnarray}
\end{mathletters}%
for a linear fit of $\DphCont/u$.

% -----------------------------------------------------------------------------

\subsection {Algorithm and Finite Volume Scaling}

Working in lattice units ($a{=}1$) with
the lattice action ($\ref{latact}$),
I update the system by heatbath and multi-grid methods
\cite{Goodman}. At each level of the multi-grid update, 
I perform over-relaxation updates.
Statistical errors are computed using the standard jackknife method.

The strategy is to vary $r$ at fixed $u$ to reach the phase transition
point.  In order to define a nominal ``phase transition'' point in
finite volume, I use the method of Binder cumulants \cite{bind}.  The Binder
cumulant of interest is defined by
\begin {equation}
   C = {\langle \bar\phi^4 \rangle \over \langle \bar\phi^2\rangle^2} ,
\end {equation}
where $\bar\phi$ is the volume average of $\phi$,
\begin {equation}
     \bar \phi \equiv {1\over V} \int d^3x \phi(x).   
\end {equation}
The nominal phase transition is defined as occurring when $C=C^*$,
where $C^*$ is a universal value which improves convergence to the
infinite volume limit.  $C^*$ depends on the shape and boundary
conditions of the total lattice volume but not on the short distance
structure.  For cubic lattice volumes with periodic boundary
conditions, $C^* \simeq 1.603(1)$ for O(1) \cite{BLHBF} and
$C^* \simeq 1.095(1)$ for O(4) \cite{MH1}. I have checked
that the errors on $C^*$ are not significant for the purposes of this
application. Therefore
I have used the central values for the nominal $C^*$.

In practice, it never happens that the simulation is done precisely at 
the $r$ where $C=C^*$. I instead simply simulate at some $r=r_{\rm sim}$
close to it.  I then use canonical reweighting of the time series data
to analyze $r$'s near $r_{\rm sim}$ and determine $r$ and
$\Delta\langle\phi^2\rangle$ at $C=C^*$.

A renormalization group analysis of the scaling of finite volume errors
shows that, when the method of Binder cumulants is used, the values of
$\Delta\phiphic$ and $\rc$ scale at large volumes as \cite{O2sim}
\begin {mathletters}
\label {eq:scaling}
\begin {equation}
   {\Delta\phiphic\over u} \simeq A_1 + A_2 (Lu)^{-d+y_t} ,
\label{eq:dphi_uafx}
\end {equation}
\begin{equation}
{r_c \over u^2} \simeq C_1 + C_2(Lu)^{-y_t-\omega} 
\label{eq:rc_uafx}
\end{equation}
\end{mathletters}%
for fixed $ua$. Here $d{=}3$ is the dimension of space,
$y_t=1/\nu$, and $\nu$ and $\omega$ are the standard O($N$)
critical exponents
associated with the correlation length and corrections to scaling,
respectively.  The values of the exponents that I have used are%
\footnote{A nice review and summary of the critical exponents can be found in \cite{Pelissetto}. For the case of O(1), we have used their summarized values of $\nu=1/y_t$ and $\omega$ based on the calculation from High Temperature(HT)
expansion technique and Monte Carlo simulations.  For comparison, some experimental results for $y_t$ are: $1.61(8)$ \cite{Sullivan}, $1.58(2)$ \cite{Chen}. For O(4), Monte Carlo
simulation for $\nu$ gives: $0.7525(10)$\cite{Ballesteros}(used by us),
$0.749(2)$\cite{Hasenbusch01}. From HT expansion:
$0.759(3)$\cite{Butera}; from $\epsilon$-expansion 0.737(8)\cite{Guida}. For $\omega$, the only MC simulation result
is $\omega=0.765$ (without quotation of error) from \cite{Hasenbusch01}. Ref.\ \cite{Guida} gives $\omega=0.774(20)$ (d=3 exp.) and 0.795(30) ($\epsilon$-exp.). We have chosen
the value to cover both of the errors for $\omega$. }
\begin{equation}
\begin{array}{lllll}
\mbox{O(1):} & \quad\quad & y_t & = & 1.587(1), \\
      & \quad\quad & \omega & = & 0.84(4), \\
\mbox{O(4):} & \quad\quad & y_t & = & 1.329(2), \\
      & \quad\quad & \omega & = & 0.79(4) .
\end{array}
\label{crit_paras}
\end{equation}
The large volume scaling of $\Delta\phiphic/u$ depends only
on $y_t$, which have errors $\sim$ 0.1\% in both cases. They have a negligible 
effect on my final results. On the other hand, for
the case of $\rc/u^2$, the large volume scaling depends on
both $y_t$ and $\omega$. I will show that the errors of 
$\omega$ do have effects on the large volume errors of 
$\rc/u^2$. They are included in the final results. 

For a discussion of higher-order terms in the large volume expansion, see
Ref.\ \cite{O2sim}, but the above will be adequate for this application.
I will fit the large $Lu$ data to the leading scaling forms
(\ref{eq:scaling}) to extract the
infinite volume limit.

The basic procedure for extracting simultaneously the $ua \to 0$ and
$Lu \to \infty$ limits of my result will be as follows.
(i) I first fix a reasonably small value of $(ua)^*$ of $ua$,
take data for a variety of sizes $Lu$ (to as large $Lu$ as practical),
and then extrapolate the size of finite
volume corrections,
fitting the coefficients $A_2$ and $C_2$ of the
scaling law (\ref{eq:scaling}).
(ii) Next, I instead fix a reasonably large physical size $(Lu)^*$ and
take data for a variety of $ua$
(to as small $ua$ as practical),
and extrapolate the continuum limit of our results at that $(Lu)^*$,
which corresponds to fitting $B_1$ and $D_1$ of (\ref{eq:Lufx1}, \ref{eq:Lufx2}).
(iii) Finally, I apply to the continuum result of step (ii) the finite
volume correction for $(Lu)^*$ determined by step (i).
In total, I have
\begin {eqnarray}
   \left\{ {\Delta\phiphic\over u} \right \}_{final}
  &=& B_1 - A_2 ((Lu)^{*})^{-d+y_t} , \\
\nonumber
   \left \{ {r_c \over u^2} \right \}_{final}
  & = & D_1 - C_2 ((Lu)^{*})^{-y_t-\omega} .
\label{fit_way1}
\end {eqnarray}

There is a finite lattice spacing error
in the extraction of the large volume correction ({\it i.e.}\ $A_2$ and
$C_2$).
In Ref.\ \cite{O2sim}, it is argued that this source of error
is formally high order in $(ua)^*$ and so expected to be small.
%I will later discuss the fit of our data to a joint dependence on
%$Lu$ and $ua$.

%%%%%%%%%%%%%%%%%%%%%%%%%%%%%%%%%%%%%%%%%%%%%%%%%%%%%%%%%%%%%%%%%%%%%%%%%%%%%

\section{Simulation Results}
\label{simresults}

\subsection {O(1) Results for {\boldmath$\Delta\phiphic/u$}}

\begin {figure}
  \vbox{
    \begin {center}
       \epsfig{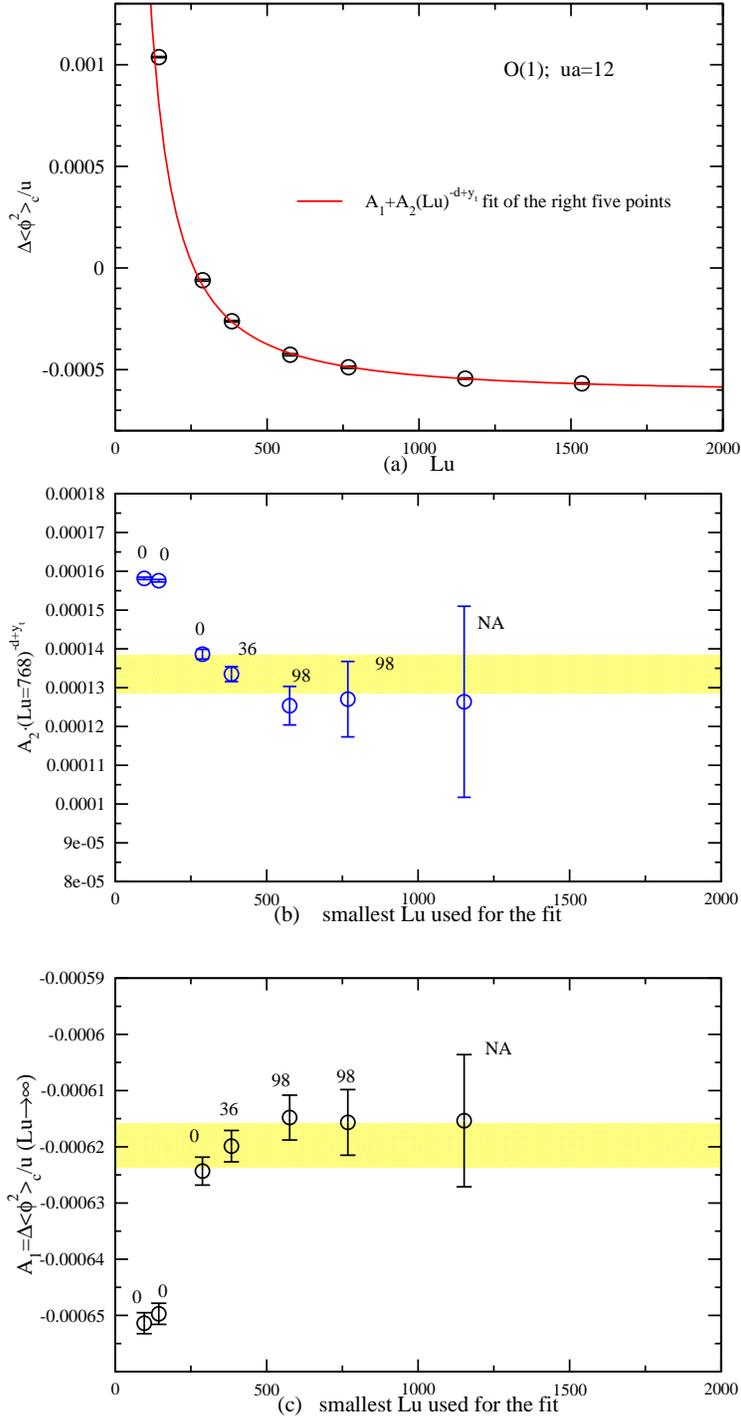}
    \end {center}
    %\vspace*{-.1in}
   \caption{
   (a) shows the simulation result for $\Delta \phiphic /u$
   vs.\ physical volume at $ua=12$ for O(1);
   (b) is the results of the difference between the $(768,12)^*$
   point and the infinite extrapolation. 
   (c) is the $Lu\rightarrow \infty$ extrapolations; and
    Since we used the
   fitting formula for $\Delta \phiphic /u$ at $(768,12)^*$
   \ref{eq:phi2renorm},
   this difference depends only on the results of $A_2$ but
   not on $A_1$. In both (b) and (c), the shaded areas are
   the quoted results. The confidence levels are given as
   as the percentage numbers on top the fitted values. ``NA''
   stands for ``non-applicable'', which means the number
   of the fitting parameters equals the number of fitted points.  
   }
   \label{u12.n1.dphi}
  }
\end {figure}

\begin {figure}
  \vbox{
    \begin {center}
       \epsfig{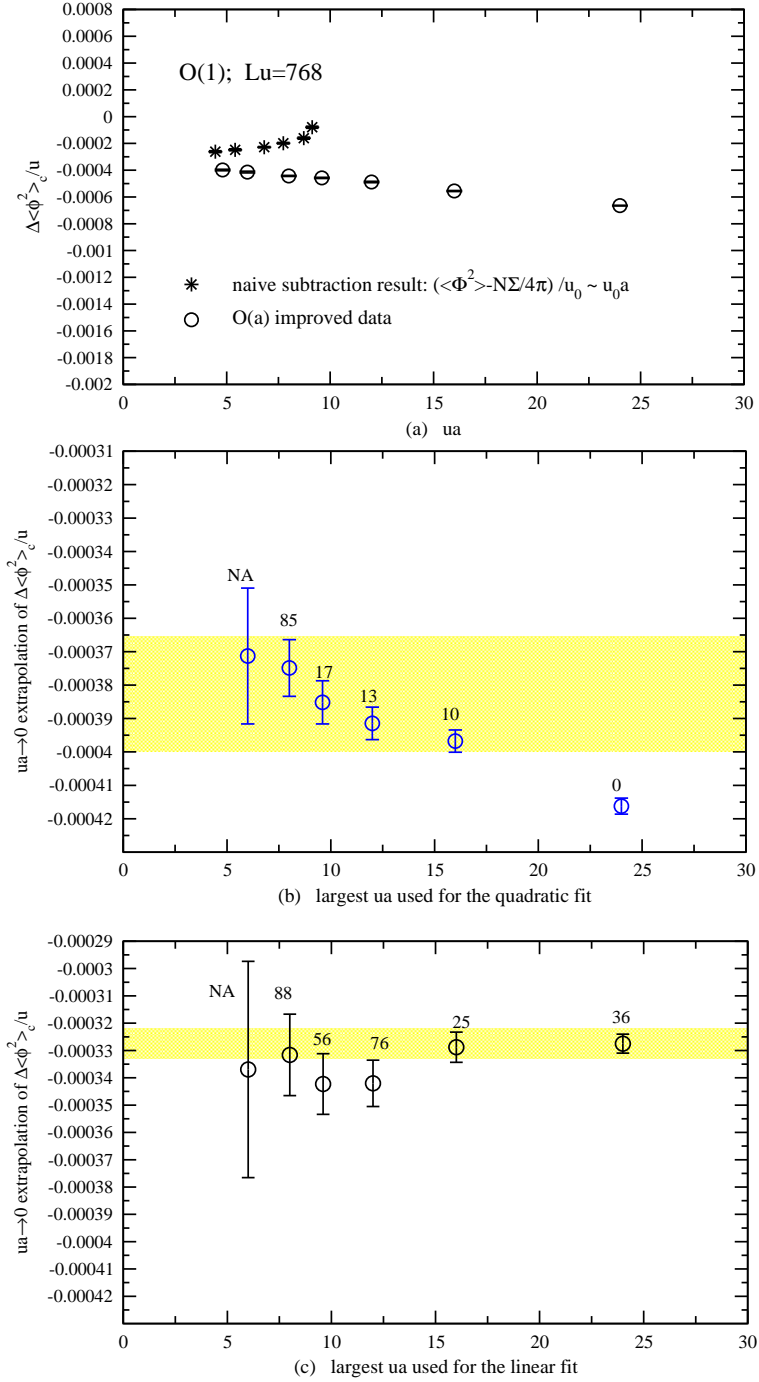}
    \end {center}
    %\vspace*{-.1in}
   \caption{Simulation results for $\Delta\phiphic/u$ 
vs $ua$ at fixed physical volume $(Lu)^*=768$ for O(1). 
The circular data are obtained from (\ref{eq:phi2renorm}), 
with $\delta \phi^2$ given by (\ref{dphi_ct}). The star
data are obtained using the naive lattice result 
$\DphLat\equiv (\langle \Phi^2 \rangle - N \Sigma/4\pi) /u_0$. 
It has linear+linear$\times$
log+quadratic dependence on $ua$. (b) gives the fitting results
for the quadratic fit and (c) gives the fitting results for
the linear fit.
%We have checked that after taking
%off the linear$\times$log term, the resulting data fits well with
%a line and extrapolates to the same continuum limit. 
}
   \label{n1.imp.768.dphi}
  }
\end {figure}

There is a practical tradeoff between how large one can take the system
size $(Lu)^*$ in order to reach the large volume limit versus how small
one can take $(ua)^*$ in order to reach the continuum limit.
Fig. \ref{u12.n1.dphi}(a) and the lower points (circles) in
\ref{n1.imp.768.dphi}(a) show,
respectively, the $Lu$ dependence at $(ua)^*=12$ and the $ua$ dependence
at $(Lu)^*=768$ for $\Delta\phiphic/u$ in the O(1) model.
From the $Lu$ dependence, we can see
that $(Lu)^*=768$ is a reasonably large value of $Lu$ --- the finite
volume corrections are moderately small.  From the $ua$ dependence, we
can see that $(ua)^*=12$ is reasonably small.

Fig.\ \ref{u12.n1.dphi}(b) shows the size $A_2 \times (Lu)^{-d-y_t}$ at
$(Lu)^*=768$ when fitting the $Lu$ dependence of Fig.\ \ref{u12.n1.dphi}(a)
to the scaling form (\ref{eq:dphi_uafx}).  The result of the
fit, and its associated confidence level, depends on how many points are
included in the fit.  Percentage confidence levels are shown in the figure.
My procedure will be to determine the best values of the fit parameters
by including as many points as possible while maintaining a reasonable
confidence level. Then to assign an error, I use the
statistical error from including one less point in the
fit. This will help avoid the underestimation of systematic errors.  The resulting estimate 0.0001335(50)
of the finite volume correction is depicted by the shaded
bar in Fig.\ \ref{u12.n1.dphi}(b) and collected in Table \ref{tab:results}.
The corresponding best fit is shown as the solid line in
Fig.\ \ref{u12.n1.dphi}(a). Even though there is no direct use of it, I show 
for completeness the fit parameter $A_1$ in Fig.\ \ref{u12.n1.dphi}(c), which corresponds
to the infinite-volume value of $\Delta\phiphic/u$ at $(ua)^*=12$. 

Having found the large volume extrapolated values, I 
will now move on to the $ua\rightarrow 0$ limit for 
$\DphCont/u$ at a fixed physical volume $(Lu)^*=768$.
Theoretically, from the argument of Ref.\ \cite{O2sim}, 
if one uses the $O(a)$ improved formula
(\ref{eq:phi2renorm}) to obtain the continuum approximated
$\DphCont/u$, the remaining lattice error should 
be $\sim O(ua)^2$. 

 The circular data in Fig.\ \ref{n1.imp.768.dphi}(a) 
show the $O(a)$ improved simulation results by using
Eq.\ (\ref{eq:phi2renorm}) .  First I have tried 
a quadratic fit with the form $B_1+B_2(ua)^2$ . The
fitting results for $B_1$ are given by Fig.\ 
\ref{n1.imp.768.dphi}(b). One can see while a 
three-point ($ua\leq 8$) fit 
gives an impressive C.L. of 85\%, including the 
fourth point ($ua=9.6$) reduces the C.L. to 17\%. 
However, adding two more points keeps the C.L. 
above 10\%. Due to this feature, it is hard 
 to determine the last point that should be
fit to the quadratic formula. Instead, I assigned
the value to cover all the $85\%$, $17\%$, $13\%$
and $10\%$ C.L.s which gives -0.000383(17).
The result is indicated by the shaded area. 

The poor behavior of the quadratic fit makes one
wonder if the data is actually more ``linear'' than
``quadratic.'' To check this, 
I re-fit the circular data using a linear function (\ref
{eq:Lufx2_dphi}). 
The results are shown in Fig.\ \ref{n1.imp.768.dphi}(c).
Surprisingly, I can fit all the data points with a
very good C.L.. The $ua\rightarrow 0$ extrapolated
value in this case is -0.0003275(55), which is very 
different from the quadratic fit result. The obvious
linear behavior of the data seem to indicate that there
might be some residual $O(a)$ coefficient in 
Eq.\ (\ref{eq:phi2renorm}) that has not been accounted
for. Given this uncertainty, 
I have used both the linear and quadratic extrapolated 
results, combined with the error due to
the finite lattice size to obtain two continuum values (Table 
\ref{tab:results}). 
They differ by about 10\%, which is considered  my
systematic error. The final result is assigned to cover 
both results and is tabulated in Tables \ref{tab:results}
and \ref{tab:summary}.%
\footnote{In Ref.\ \cite{O2sim}, for the case of 
O(2) model, the $Lu$ fixed data
are fitted by quadratic functions only. The $Lu=144$
data can be fitted also by a linear function. The
$Lu=576$ data on the other hand can be fitted by 
only a quadratic function for the first four points.
However if one takes off the smallest $ua$ point, the
rest four points can be fitted very well by a 
straight line too. This might indicate that there
could be also some linear coefficients in the O(2)
theory. For more discussions on this possible
linear coefficient see Ref.\ \cite{thesis}.} 

As a comparison, I have also shown the naive subtracted
result given by 
$\DphLat/u_0 \equiv (\<\Phi^2\>-N\Sigma/4\pi)/u_0$ vs.\ 
the unimproved $u_0a$ as the scale based
on the same simulation (the stars in Fig.\ \ref{n1.imp.768.dphi}(a)).  
%Theoretically, $\DphLat/u_0$ should have $log\times linear$
%and $linear$ dependences on $u_0a$. This then should show 
%a more explicit effects of lattice spacing errors. 
One can see that the finite lattice spacing errors is more 
explicit with the stars, corresponding to the 
unimproved data, than that of the circles, corresponding
to the improved ones (using (\ref{eq:phi2renorm})). Theoretically
 $\DphLat/u_0$ data have $log\times linear$ and $linear$ dependences 
on $u_0a$.
%  However
% the linear fit for the first four star points gives
%  $-0.000347(16)$ with 19\% C.L.. This differs from
%  the quadratic fit by about $10\%$, which is considered
% as my systematic error. Of course, one
% might doubt the accuracy of this linear extrapolation, 
% since the dependence on $ua$ contains a $log$ piece, which
% is hard to control.%

\subsection {{\boldmath$\rc$} and O(4) Results}
\begin {figure}
  \vbox{
    \begin {center}
       \epsfig{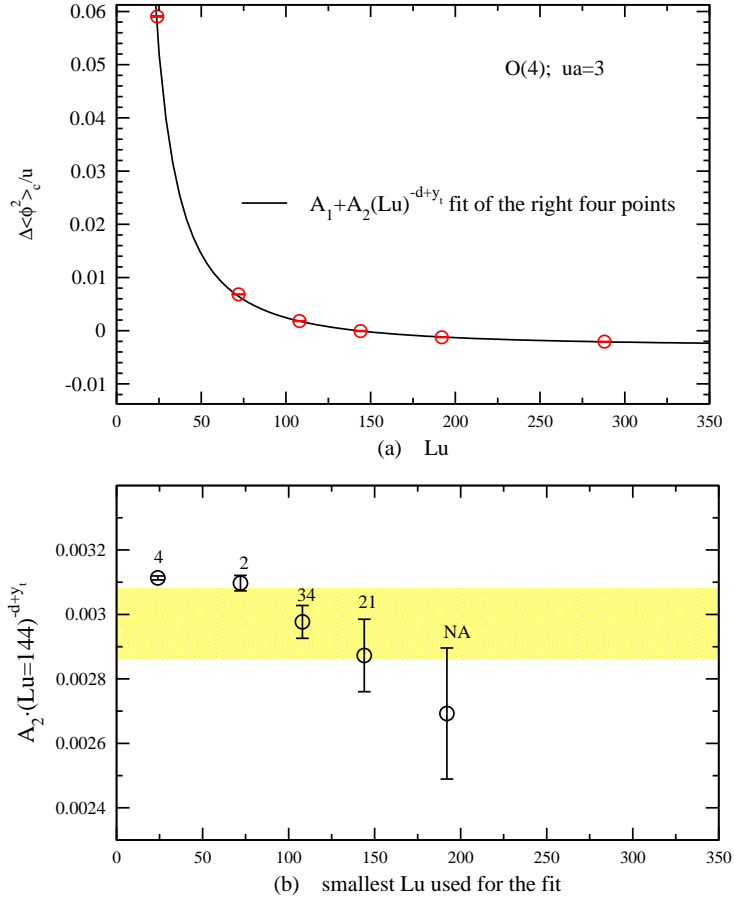}
    \end {center}
    %\vspace*{-.1in}
   \caption{As Fig.\ \ref{u12.n1.dphi} but for O(4)
at $ua=3$ with reference point $(Lu,ua)^*=(144,3)$. 
}
   \label{u3.n4.dphi}
  }
\end {figure}

\begin {figure}
  \vbox{
    \begin {center}
       \epsfig{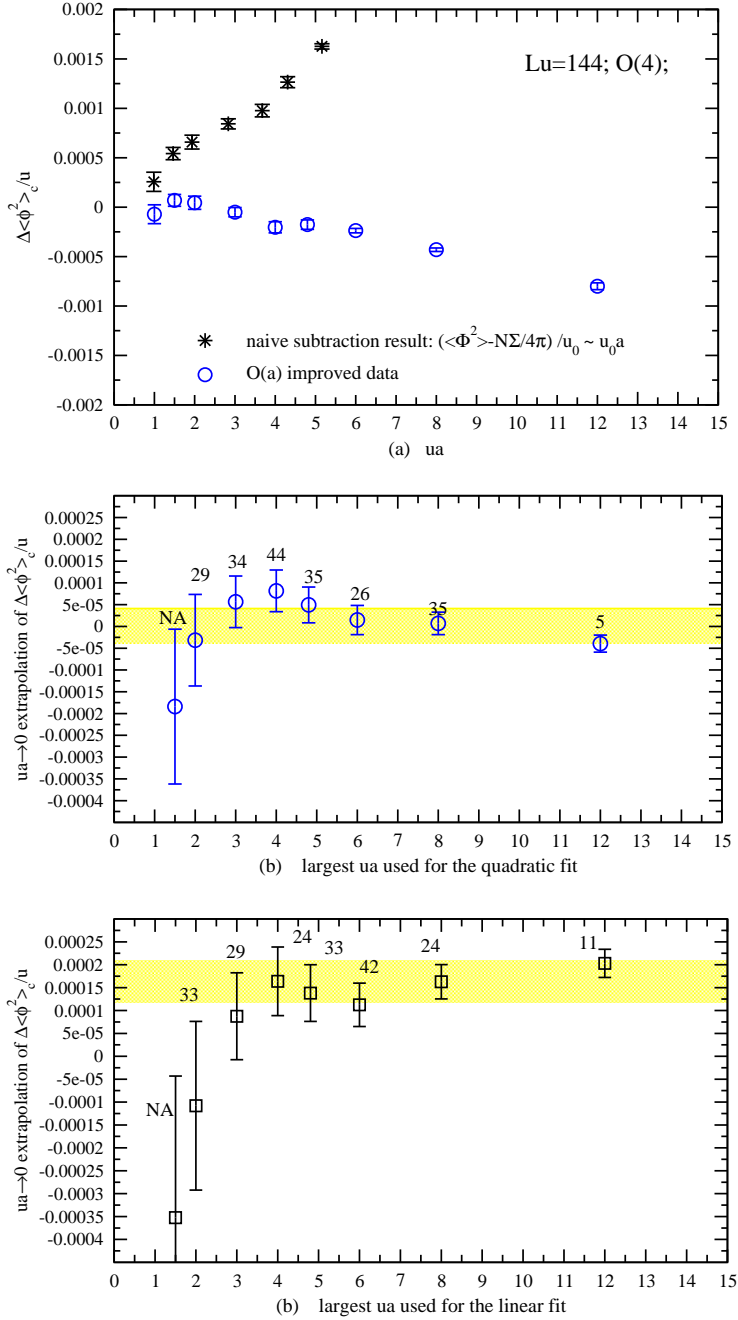}
    \end {center}
    %\vspace*{-.1in}
   \caption{As Fig.\ \ref{n1.imp.768.dphi}, but for O(4) at $(Lu)^*=144$. The triangular data in (a) is again the naive lattice result $\DphLat/u_0$ which shows again bigger lattice errors than the $O(a)$ improved data(circular data).  
}
   \label{n4.imp.144.dphi}
  }
\end {figure}

Figs.\ \ref{u3.n4.dphi} and \ref{n4.imp.144.dphi}
show similar curves for the
O(4) model, where we have taken $(Lu)^*=144$ and $(ua)^*=3$ as
a reasonably large size and reasonably small lattice spacing.
%One can check from the figures that the corrections at these values
%are moderately small.
It's worth noting the large $N$ limit predicts that the
distance scale that characterizes the physics of interest should
scale as $1/(Nu)$.  That leads one to
expect that the upper limit for ``reasonably small'' values of $ua$ 
and lower limit for ``reasonably large'' values of $Lu$ should scale
roughly as $1/N$.

The fitting proceeds as in the O(1) case. For the
$Lu$-fixed data of $\DphCont/u$ (Fig.\ \ref{n4.imp.144.dphi}), 
I have again fitted the data with both linear
and quadratic functions. Both the quadratic fit and
the linear fit can go up to $ua=8$ while keeping 
good C.L.s. However their $ua\rightarrow 0$ extrapolations 
are also different (see Fig.\ \ref{n4.imp.144.dphi} and 
Table \ref{tab:results} for the fitting results). The final
continuum value is assigned to cover both extrapolations.

\begin {figure}
  \vbox{
    \begin {center}
       \epsfig{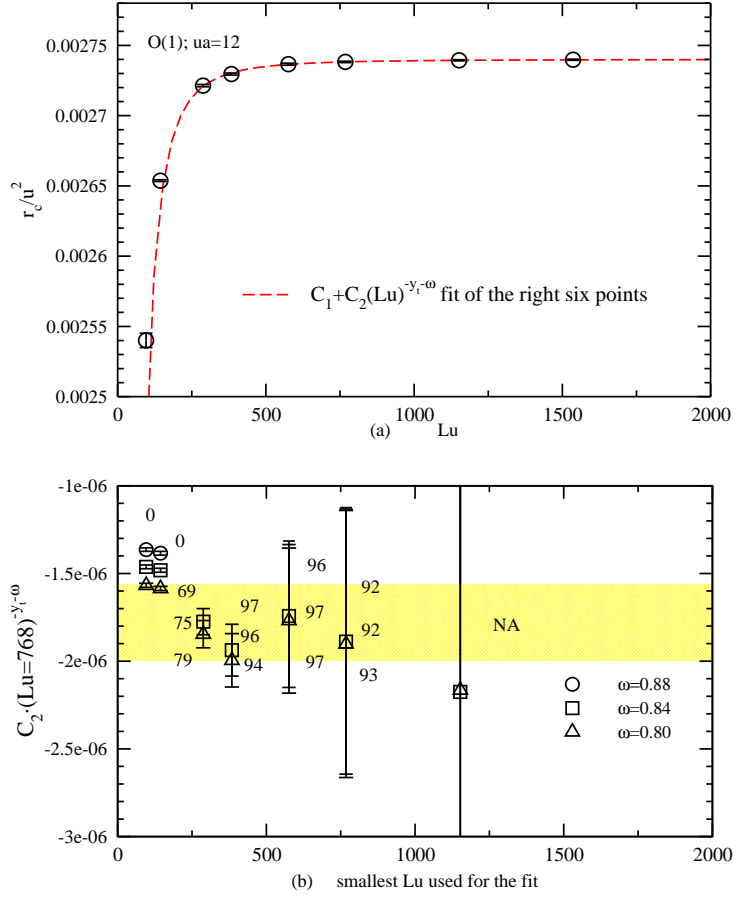}
    \end {center}
    %\vspace*{-.1in}
   \caption{Simulation results for $r_c /u^2$ vs physical
volume for O(1) at $ua=12$ with $(Lu,ua)^*=(768,12)$. We have used three values for $\omega$ for the fits. The confidence levels for all the fits are listed
in the same vertical order as the legend. The final assignment covers all the three extrapolations.  The changes of the extrapolated $C_2(Lu)^{-y_t+\omega}$ is about one error bar due to the uncertainty in $\omega$. 
}
   \label{u12.n1.imp.rc}
  }
\end {figure}

\begin {figure}
  \vbox{
    \begin {center}
       \epsfig{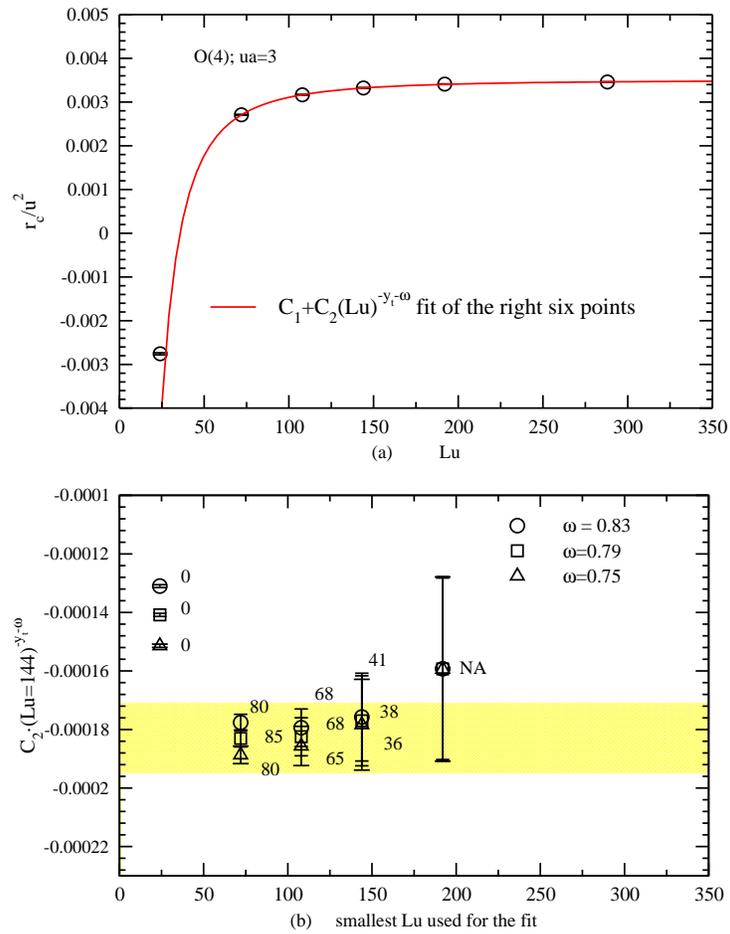}
    \end {center}
    %\vspace*{-.1in}
   \caption{As Fig.\ \ref{u12.n1.imp.rc} but for O(4) at 
$ua=3$ with $(Lu,ua)^*=(144,3)$.
}
   \label{u3.n4.rc}
  }
\end {figure}

\begin {figure}
  \vbox{
    \begin {center}
       \epsfig{file=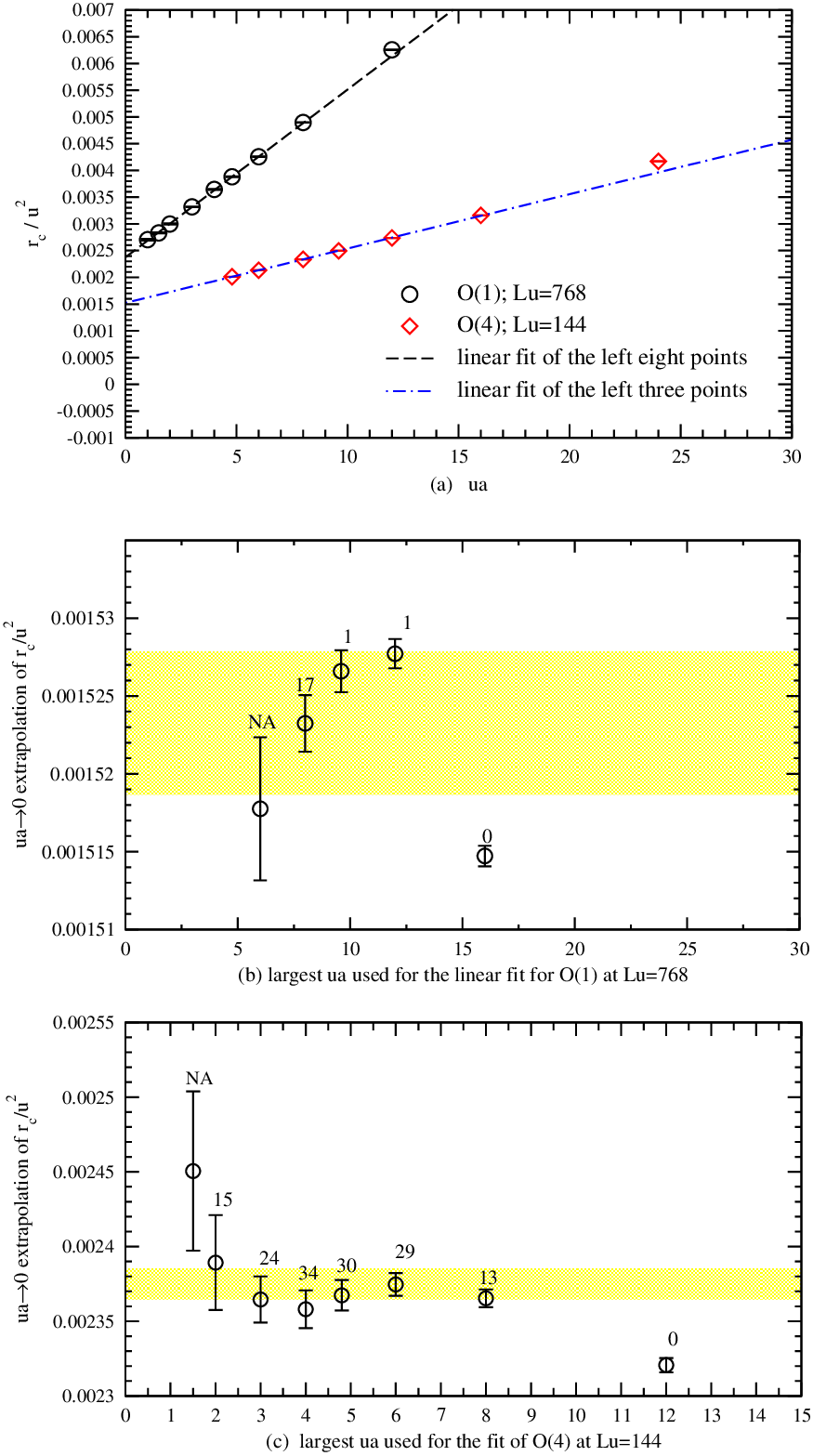,scale=.7,angle=0}
    \end {center}
    %\vspace*{-.1in}
   \caption{Simulation results for $r_c/u^2$ vs $ua$ 
for O(1) at $(Lu)^*=768$ and for O(4) at $(Lu)^*=144$.
The two straight lines in (a) are linear fits to the data. 
}
   \label{n1.n4.rc}
  }
\end {figure}

The analysis of the result for $\rc/u^2$ in both the O(1) and 
O(4) models is much the same 
(Figs.\ \ref{u12.n1.imp.rc}--\ref{n1.n4.rc}).  
For the large volume extrapolations, since the critical
exponent $\omega$ has a larger uncertainty ($\sim 5\%)$, 
I have used three different values of $\omega$ 
(maximum, central, minimum) for both O(1) and O(4)'s cases. 
The final results of the $C_2\times(Lu)^{-y_t-\omega}$ value 
cover all the three different extrapolated values. 
This difference results in about $1\%$ of the error quoted
in the final results for $\rc/u^2$. For the $Lu$ fixed
data, since there is no improvement of $O(a)$ errors, the
 continuum extrapolation of the data 
taken at fixed $Lu$ are fitted to linear behavior in $ua$, 
following (\ref{eq:Lufx1_rc} or \ref{eq:Lufx2_rc}). 
The fitting results are again summarized in Tables \ref{tab:summary} 
and \ref{tab:results}.
  
%=====================
%-------------------
%=====================
%-------------------
%=====================
%---------------------
%-------------------
%=====================

% %%%%%%%%%%%%%%%%%%%%%%%%%%%%%%%%%%%%%%%%%%%%%%%%%%
% \begin{table}
% \begin{center}
% \begin{tabular} {|c|c|c|c|c|c|l|}
% \hline
% \multicolumn{7}{|c|}{${\Delta\phiphic/ u}$} \\
% \hline
% N & $(Lu,ua)^*$
%   & \multicolumn{3}{c|}{$A_1+A_2(Lu)^{-d+y_t}$ fit }
%   & \multicolumn{1}{c|}{$B_1+B_2(ua)^2$ fit}
%   & \multicolumn{1}{c|}{${\Delta\phiphic /  u} =$} 
% \\ \cline{3-6}   
%  & & $y_t$ & $A_1$ & $A_2(Lu)^{-d+y_t}$ & $B_1$ & $B_1-A_2(Lu)^{-d+y_t}$ \\
% \hline
%  1 &(768,12) & 1.587(1) &-0.0006199(40)& 0.0001335(51) & -0.0003968(49) & -0.0005250(82)\\
% \cline {1-7} 
% \cline{2-3} \cline{5-7}
% 4 & (144,3) & 1.329(2) &-0.003023(68) &0.00297(11) & 0.000047(49)  & -0.00292(12)\\
% \hline \hline
% \multicolumn{7}{|c|}{${r/ u^2}$} \\
% \hline
% N & $(Lu,ua)^*$
%   & \multicolumn{3}{c|}{$C_1+C_2(Lu)^{-y_t-\omega}$ fit }
%   & \multicolumn{1}{c|}{$D_1+D_2(ua)$ fit}
%   & \multicolumn{1}{c|}{$r_c / u^2 =$} 
% \\ \cline{3-6}   
%  & & $\omega$ & $C_1$ & $C_2(Lu)^{-y_t-\omega}$ & $D_1$ & $D_1-C_2(Lu)^{-y_t-\omega}$ \\
% \hline
%  1 & (768,12) & 0.84(4) & 0.00274001(28)& -0.00000178(22)
%  & 0.0015231(48)  & 0.0015249(48) \\
% \hline
%  4 & (144,3) & 0.79(4) &0.003505(10) &-0.000183(12)& 
%  0.002368(13)  & 0.002551(18) \\
% \hline
% \end{tabular}
% \end{center}
% \caption{Fitting results for $\Delta\phiphic /u$ and $r_c/
% u^2$ using formula \ref{fit_way1}}.
% \label{tab:results}
% \end{table}
% %%%%%%%%%%%%%%%%%%%%%%%%%%%%%%%%%%%%%%%%%%%%%%%%%%
\begin{landscape}
 \begin{table}
 \begin{center}
 \begin{tabular}{|c|c||c|c||c|c||l|l|l|}
 \hline
 \multicolumn{9}{|c|}
  {\rule[-2mm]{0mm}{6mm} $\Delta\phiphic/ u$} \\
 \hline
  {\rule[3mm]{0mm}{5mm} $N$} 
   & $(Lu,ua)^*$
   & \multicolumn{2}{c||}
   {\rule[-3mm]{0mm}{3mm}$A_1+A_2\times (Lu)^{-d+y_t}$ fit}
   & \multicolumn{1}{c|}
   {$B_1+B_2(ua)^2$ fit } 
   & \multicolumn{1}{c||} 
   {$B_1+B'_2(ua)$ fit } 
   & \multicolumn{3}{c|}
   {$\Delta\phiphic /u=B_1-A_2 \times (Lu)^{-d+y_t}$} 
\\ \cline{3-9}
  & 
  & {\rule[1mm]{0mm}{4mm} $A_1$} 
   & $A_2 \times (Lu)^{-d+y_t}$ 
   & \multicolumn{2}{c||}{ $B_1$} 
   &  \multicolumn{1}{c|}{$O(ua)^2$} 
  &  \multicolumn{1}{c|}{$O(ua)$} 
 &\multicolumn{1}{c|}{final}\\
 \hline
  {\rule[-2mm]{0mm}{6mm} 1} 
  & { (768,12)} & -0.0006199(40) & 0.0001335(50) & -0.000383(17) & -0.003275(55) & -0.000517(18) & -0.0004610(74)
  & -0.000494(41)
 \\ \hline
 {\rule[-2mm]{0mm}{6mm} 4}
  & (144,3)  &-0.003023(92) &0.002978(92) & 0.000007(26)
  & 0.000163(37)  & -0.002971(96) & -0.002815(99)
  & -0.00289(18)
  \\  \hline \hline
 \multicolumn{9}{|c|}{\rule[-2mm]{0mm}{6mm} $\rc/u^2$} 
 \\ \hline
  {\rule[3mm]{0mm}{5mm} $N$} 
   & $(Lu,ua)^*$
   & \multicolumn{2}{c||}{
 \rule[-3mm]{0mm}{3mm} $C_1+C_2\times (Lu)^{-y_t-\omega}$ fit }
   & \multicolumn{2}{c||}{
  $D_1+D_2(ua)$ fit}
   & \multicolumn{3}{c|}
  {
 %\rule[3mm]{0mm}{3mm} 
   $r_c / u^2=D_1-C_2 \times (Lu)^{-y_t-\omega}$} 
 \\ \cline{3-6}   
  & 
  & {\rule[1mm]{0mm}{4mm} $C_1$}
   & $C_2\times (Lu)^{-y_t-\omega}$ &
  \multicolumn{2}{c||} {$D_1$} 
  & \multicolumn{3}{c|} {} \\
 \hline
    {\rule[-2mm]{0mm}{6mm} 1} 
  & (768,12) & 0.00274001(28)& -0.00000178(22)
  & \multicolumn{2}{c||}{0.0015231(48)}
    & \multicolumn{3}{c|}{0.0015249(48)} \\
 \hline
    {\rule[-2mm]{0mm}{6mm} 4} 
  & (144,3)  
  & 0.0035035(62) 
  &-0.000183(12)
  & \multicolumn{2}{c||}{0.002375(10)}
   &\multicolumn{3}{c|}{0.002558(16)}\\
 \hline
 \end{tabular}
 \end{center}
 \vspace{2cm}
 \caption{
 Fitting results for $\Delta\phiphic /u$ and $\rc/
 u^2$ using formula (\ref{fit_way1}). In the case of 
$\DphCont/u$, both a linear fit result and a quadratic
fit result for $B_1$ are given. The final values are
 assigned to cover both fit results.  }
\label{tab:results}
\end{table}
\end{landscape}

\section{Conclusion}
\label{concls}

\begin {figure}
  \vbox{
    \begin {center}
       \epsfig{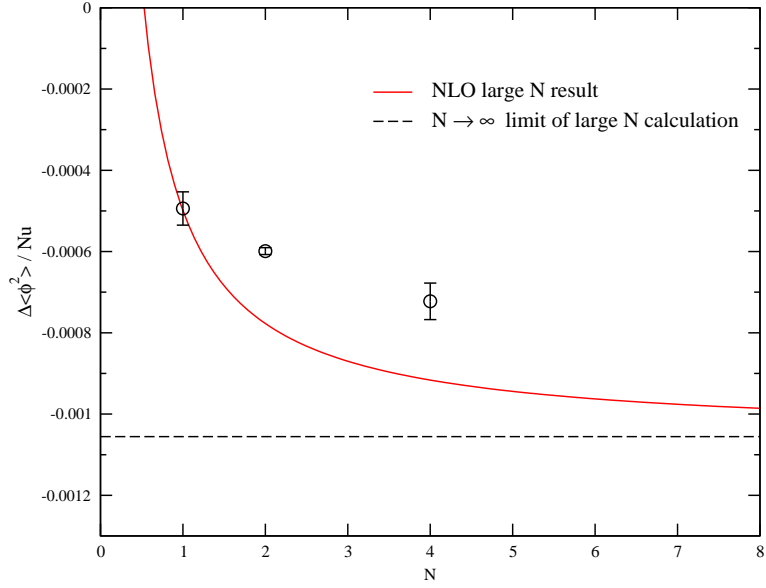}
    \end {center}
    %\vspace*{-.1in}
   \caption{Comparison of NLO large $N$ result with
   simulation results for O(1), O(2) and O(4).
   The solid line is the NLO large $N$ result for $\Delta\phiphic/Nu$,
   and the dashed line is the $N{=}\infty$ value.
   }
   \label{n1.n2.n4.dphi}
  }
\end {figure}

Fig.\ \ref{n1.n2.n4.dphi} and Table \ref{tab:results} show
a comparison
of the simulation results for $\Delta\phiphic/u$
with the NLO result (\ref{eq:dphiN}) for
$N=1,2,4$. The $N=1,4$ cases are from this paper.
%Due to the systematic uncertainty in the $ua\rightarrow
%0$ extrapolation, they have relatively bigger errors.
The result of O(2) model is taken from Ref.\ \cite{O2sim}.%
\footnote{If one also considers the systematic uncertainty
in the $ua\rightarrow 0$ extrapolation in Ref.\ \cite{O2sim}, then 
the result for O(2) should have a higher 
systematic uncertainty.} 
In the figure, I have actually plotted $\Delta\phiphic/(Nu)$, where
the explicit factor of $N$ factors out the leading-order dependence on $N$ as $N\to\infty$.
Amusingly, the large $N$ estimate for $N{=}1$ seems to be
 the most accurate of the three cases.  This is presumably 
an accident---the result of the large $N$ approximation just happens to be crossing the set of
actual values near $N=1$. From Eq. (\ref{eq:dphiN}), 
the error of the for $\Delta \phiphic / Nu $ should scale as $1/N^2$, but there is
clearly no sign of such behavior between $N{=}2$ and $N{=}4$. This might indicate that $N{=}4$ is still too small for the error in the large $N$
approximation to scale properly. It would be interesting to numerically
study yet higher $N$ such as $N{=}8$ or $N{=}16$ to attempt to verify the details of the approach to the large $N$ limit. 

For the systematic error due to the possible linear 
behavior in the $ua\rightarrow 0$ extrapolation of
$\DphCont/u$, 
so far no theoretical reasoning is found. It 
would be interesting to reinvestigate the matching 
calculations given in Ref.\ \cite{O2sim}.
% in order
%to obtain the $O(a)$ improved formula (\ref{eq:phi2renorm}).

%\section{Acknowledgements}
\acknowledgements
I would like to thank the physics department of the 
University of  Virginia for providing me the opportunity 
and support for this research work. The work was 
supported, in part, by the U.S.\ Department of Energy 
under Grant No.\ DE-FG02-97ER41027.  I am especially 
grateful to P.\ Arnold for his valuable guidance and 
assistance throughout the research work.

% With the method developed in Ref.\ \cite{O2sim} and used in
% this paper, this would first require a determination of 
% the critical value of the Binder cumulant for those 
% theories, which I have not attempted. 

%---------------------
%-------------------

\clearpage
\appendix
\section{Tabulated Data}
\label{data.table}
Tables \ref{O1data} and \ref{O4data} are a collection
of all the data reported in this paper. The standard 
integrated decorrelation time $\tau$ for a single 
operator $O$ is defined
\begin {equation}
    \tau \equiv {1\over2} + \sum_{n=1}^\infty {C(n)\over C(0)} ,
 \label {eq:tau}
\end {equation}
where
\begin {equation}
   C(n) =
      {1\over(N-n)} \sum\limits_{i=1}^{N-n} A_i A_{i+n}
      - \left({1\over N}\sum\limits_{i=1}^N A_i\right)^2 ,
 \end {equation}
 is the auto-correlation function associated with the operator $A$. In practice the sum in (\ref{eq:tau}) is cut-off 
when $C(n)/C(0)$ first drops below $0.05$ because of 
the statistical fluctuation in $C(n)$. The nominal
decorrelation time listed in Tables \ref{O1data} and \ref{O4data} is the largest value of the various operators
required in the computations of the Binder cumulant
and $\DphCont/u$ by using the canonical reweighting
method (\cite{O2sim}).
% statistics. Our nominal decorrelation time is the largest integrated decorrelation time
% of the various expectations required in the computation of the Binder cumulant
% and of $\langle\phi^2\rangle$ (See Ref.\ \cite{O2sim} for the definitions of those operators).

\begin{table}
\begin {center}
\setlength{\tabcolsep}{8pt}
\begin {tabular}{rlrllcr}   
  \\  \hline\hline 
\multicolumn{7}{c}{O(1)} 
\\                      \hline
$Lu$ & $L/a$ & $ua$ & \multicolumn{1}{c}{$r_\c/u^2$}
  & \multicolumn{1}{c}{$\Delta\phiphic/u$}
  & $\tau_{\rm decorr}$
  & $N_{\rm sweeps}/2\tau$
  \\  \hline
 8. & ~8 &1. &~-0.01975(18) &~0.14599(50) &  0.8 & 61617\\
 12. & ~12 &1. &~-0.00901(12) &~0.07718(41) &  0.9 & 58809\\
 24. & ~8 &3. &~-0.001194(44) &~0.02577(14) &  1.1 & 47134\\
 36. & ~12 &3. &~0.000437(31) &~0.01332(11) &  1.1 & 46288\\
 48. & ~16 &3. &~0.001059(10) &~0.008149(54) &  1.1 & 44135\\
 96. & ~8 &12. &~0.0025401(52) &~0.002530(23) &  2.5 & 20188\\
 144. & ~12 &12. &~0.00265373(60) &~0.0010371(26) &  2.3 & 590789\\
 192. & ~32 &6. &~0.0020919(35) &~0.000446(20) &  3.5 & 7175\\
 288. & ~24 &12. &~0.00272132(81) &~-0.0000604(44) &  4.3 & 53716\\
 384. & ~32 &12. &~0.00272961(71) &~-0.0002622(36) &  5.8 & 19281\\
 576. & ~48 &12. &~0.00273662(70) &~-0.0004271(50) & 11.3 & 6731\\
 768. & ~128 &6. &~0.00213384(75) &~-0.0004142(63) &  $1.6^*$ & 4795\\
 768. & ~160 &4.8 &~0.00201051(74) &~-0.0003987(61) &  $2.8^*$ & 1643\\
 768. & ~32 &24. &~0.00416985(16) &~-0.0006645(23) &  3.7 & 13690\\
 768. & ~48 &16. &~0.00315905(45) &~-0.0005548(39) & 11.5 & 8800\\
 768. & ~64 &12. &~0.00273822(48) &~-0.0004886(51) &  8.9 & 22219\\
 768. & ~80 &9.6 &~0.00249587(61) &~-0.0004567(42) &  $1.3^*$\footnotemark & 4545\\
 768. & ~96 &8. &~0.00233612(56) &~-0.0004429(39) &  $2.7^*$ & 15260\\
 1152. & ~96 &12. &~0.00273936(42) &~-0.0005441(33) &  $3.7^*$ & 3480\\
 1536. & ~128 &12. &~0.00273977(41) &~-0.0005679(33) & $7.5^*$ & 1411\\

\hline
\end {tabular}
\end {center}
\caption{Collection of O(1) simulation data}
\label{O1data}
\end {table}
\footnotetext{Numbers with $*$ means the simulation data  are collected every 10$\tau_{decorr}$ sweeps.} 

\begin{table}
\begin {center}
\setlength{\tabcolsep}{8pt}
\begin {tabular}{rlrllcr}   
\\\hline\hline
\multicolumn{7}{c}{O(4)} 
\\ \hline
$Lu$ & $L/a$ & $ua$ & \multicolumn{1}{c}{$r_\c/u^2$}
  & \multicolumn{1}{c}{$\Delta\phiphic/u$}
  & $\tau_{\rm decorr}$
  & $N_{\rm sweeps}/2\tau$
  \\  \hline
 4. & ~4 &1. &~-0.13588(100) &~1.1014(54) &  0.9 & 53759\\
 8. & ~4 &2. &~-0.04051(27) &~0.3673(17) &  0.9 & 112355\\
 12. & ~4 &3. &~-0.0182(17) &~0.193(13) &  0.5 & 100000\\
 16. & ~4 &4. &~-0.00913(11) &~0.11794(55) &  1.6 & 60697\\
 24. & ~8 &3. &~-0.002756(22) &~0.05904(11) &  1.1 & 465111\\
 32. & ~16 &2. &~-0.000403(54) &~0.03485(29) &  1.0 & 25913\\
 48. & ~16 &3. &~0.001792(22) &~0.01641(13) &  1.1 & 44839\\
 72. & ~24 &3. &~0.002709(11) &~0.006827(59) &  1.4 & 58782\\
 108. & ~36 &3. &~0.0031673(87) &~0.001814(54) &  1.9 & 26085\\
 144. & ~12 &12. &~0.0062556(51) &~-0.0008(34) &  1.7 & 57244\\
 144. & ~144 &1. &~0.002704(17) &~-0.000072(95) &  4.6 & 4416\\
 144. & ~18 &8. &~0.0048967(29) &~-0.000431(18) &  2.3 & 154272\\
 144. & ~24 &6. &~0.0042556(34) &~-0.000238(22) &  2.3 & 83972\\
 144. & ~30 &4.8 &~0.0038808(76) &~-0.000177(48) &  2.8 & 18519\\
 144. & ~36 &4. &~0.0036449(94) &~-0.000204(57) &  2.5 & 47481\\
 144. & ~48 &3. &~0.003318(66) &~-0.000051(47) &  1.6 & 47051\\
 144. & ~72 &2. &~0.002998(12) &~0.000045(67) &  3.3 & 8952\\
 144. & ~8 &18. &~0.0087128(29) &~-0.000018(41) &  1.1 & 44435\\
 144. & ~9 &16. &~0.0077983(67) &~-0.000689(62) &  1.0 & 452306\\
 144. & ~96 &1.5 &~0.0028309(98) &~0.000069(60) &  3.2 & 11499\\
 192. & ~64 &3. &~0.0034097(76) &~-0.001234(46) &  2.8 & 15893\\
 288. & ~96 &3. &~0.0034596(63) &~-0.002053(42) &  7.7 & 5792\\

\hline
\end {tabular}
\end {center}
\caption{Collection of O(4) simulation data}
\label{O4data}
\end {table}
\bigskip

%-------------------
%%%%%%%%%%%%%%%%%%%
%REFERENCE
%%%%%%%%%%%%

%%%%%%%%%%%%%%%%%%%%%%%%%%%%%%%%%%%%%%%%%%%%%%%%%%

\end{document}